\documentstyle[twocolumn,prl,aps]{revtex}

\begin{document}
\draft

\catcode`=11 \catcode`=12 
\twocolumn[\hsize\textwidth\columnwidth\hsize\csname@twocolumnfalse\endcsname
\title{Exact soliton solution and inelastic two-soliton collision
 in spin chain driven by a time-dependent magnetic field   }
\author{ Zai-Dong Li$^1$, Lu Li$^1$,  W. M. Liu$^{2,3}$,
Jiu-Qing Liang$^1$, T. Ziman$^3$ }
\address{$^1$Institute of Theoretical Physics and Department of Physics,
Shanxi University, Taiyuan 030006, China} 
\address{$^2$National Lab of Magnetism, Institute of Physics, Chinese 
Academy of Sciences, Beijing 100080, China}
\address{$^3$Institut Laue Langevin, Grenoble 38042, France}

\date{\today}
\maketitle

\begin{abstract}
We investigate dynamics of exact N-soliton trains in spin chain driven by a time-dependent magnetic field
 by means of an inverse scattering transformation. The one-soliton 
solution indicates obviously the spin precession around the 
magnetic field and periodic shape-variation induced by the time varying field as well.
 In terms of the general soliton solutions N-soliton interaction and particularly various two-soliton
 collisions are analyzed. The inelastic collision by which we mean the soliton shape
 change before and after collision appears generally due to the time varying field. 
We, moreover, show that complete inelastic collisions can be achieved by adjusting 
spectrum and field parameters. This may lead a potential technique of shape control 
of soliton.
\end{abstract}

\pacs{PACS numbers:  05.90.+m, 04.20.Jb, 05.45.Yv, 75.10.Hk}]

\section{Introduction}

Over the past three decades, an enormous amount of literature has appeared
throughout soliton physics and the underlying completely integrable models.
The classical Heisenberg spin chain which exhibits both coherent and chaotic
structures depending on the nature of the magnetic interactions\cite
{Trullinger,Kivshar,Ablowitz,Fokas} has attracted considerable attentions in
nonlinear science and condensed-matter physics. Solitons in quasi
one-dimensional magnetic systems have already been probed experimentally by
neutron inelastic scattering \cite{Kjems78,Boucher90}, nuclear magnetic
resonance \cite{Boucher80,Jongh81}, Mossbauer linewidth measurements \cite
{Thiel81}, and electron spin resonance \cite{Asano00}. The corresponding
theoretical studies are based usually on the Landau-Lifshitz equation \cite
{landau}. The isotropic spin chain has been studied in various aspects\cite
{Laksmanan,Fogedby,Takhtajan,Shimizu,Tjon} and the construction of soliton
solutions of Landau-Lifschitz equation with an easy axis has been also
discussed\cite{Bolovik,Quispel}. It is demonstrated that the inverse
scattering transformation \cite{Takhtajan,Borovik,Boro,Chen} can be used to
solve the Landau-Lifschitz equation for an anisotropic spin chain. Great
efforts \cite{Mikeska,Long} have been devoted to construct the soliton
solution which are found by means of the Darboux transformation \cite{Huang}%
. The continuum spin chain in an external magnetic field is of great
interest and multi-soliton solutions of Landau-Lifschitz equation for an
isotropic spin chain have been reported \cite{Pu}. Using Darboux
transformation the nonlinear dynamics of anisotropic Heisenberg spin chain
in an external magnetic field is investigated and exact soliton solutions
are obtained\cite{Liu}. Recently soliton interaction has been investigated%
\cite{Tjon}. The main goal of this paper is to study the new effect of
soliton-soliton interaction in spin chain driven by time oscillating
magnetic field. We obtain exact solution of N-soliton trains in terms of an
inverse scattering transformation. It is shown that inelastic collisions
generally appear due to the time-varying field and the complete inelastic
collisions which may lead to a interesting technique of soliton filter and
switch can be achieved in special case.

The outline of this paper is organized as follows: In Sec. II the formalism
obtained by an inverse scattering transformation is explained in detail and
the general N-soliton solution for reflectionless case is obtained.
Precession of nonlinear spin waves in the oscillating magnetic field is
shown in Sec. III. Sec. IV is devoted to general two-soliton solution and
soliton collisions. Finally, Sec. V will give our concluding remarks.

\section{Exact solution of N-soliton train}

Our starting Hamiltonian describing the spin chain in a time oscillating
magnetic field with an arbitrary direction can be written as 
\begin{equation}
\hat{H}=-J\sum_{<n,n^{\prime }>}\hat{S}_n\cdot \hat{S}_{n^{\prime }}-g\mu _B%
{\bf B}\left( t\right) \cdot \sum_n\hat{S}_n,
\end{equation}
where $\widehat{S}_n\equiv (\widehat{S}_n^x,\widehat{S}_n^y,\widehat{S}_n^z)$
with $n=1,2,...,N$ are spin operators, $J>0$ is the pair interaction
parameter, $g$ the Lande factor and $\mu _B$ is the Bohr magneton, ${\bf B}%
\left( t\right) =B\cos (\omega t){\bf e}$ is the external magnetic field
with ${\bf e}=\left( \sin \theta ,0,\cos \theta \right) $ denoting the unit
vector of field direction where chain axis and direction of magnetic field
are assumed in x-z plane. The angle $\theta $ between direction of magnetic
field and z-axis is arbitrary.

The equation of motion for the spin operator on the nth site is $\frac d{dt}%
\hat{S}_n=-\frac i\hbar [\hat{S}_n,\hat{H}]$. At low temperature, the spin
can be treated as a classical vector such that $\hat{S}_n\rightarrow {\bf S}%
(x)$. So that the equation of motion in a continuum spin chain under a
time-dependent magnetic field can be obtained as a Landau-Lifschitz type 
\begin{equation}
\frac \partial {\partial t}{\bf S}={\bf S}\times \left( \frac{\partial ^2}{%
\partial x^2}{\bf S}+{\bf \varepsilon }\right) ,  \label{Eq2}
\end{equation}
with ${\bf \varepsilon }=g\mu _B{\bf B}\left( t\right) /\left( 2J\right) $,
where ${\bf S}(x,t)=(S^x(x,t)$, $S^y(x,t)$, $S^z(x,t))$. We set the length
of the spin vector to unit for the sake of simplicity ${\bf S}^2\left(
x,t\right) =1$. The dimensionless time $t$ and coordinate $x$ in Eq. (\ref
{Eq2}) are scaled in unit $\frac 1{2J}$ and $d$ respectively, where $d$
denotes the lattice constant.

The corresponding Lax equations for the equation of motion (\ref{Eq2}) are
written as 
\begin{eqnarray}
\frac \partial {\partial x}\Psi \left( x,t,\lambda \right) &=&L\left(
\lambda \right) \Psi \left( x,t,\lambda \right) ,  \nonumber \\
\frac \partial {\partial t}\Psi \left( x,t,\lambda \right) &=&M\left(
\lambda \right) \Psi (x,t,\lambda ),  \label{lax1}
\end{eqnarray}
where $\lambda $ is the spectral parameter, $\Psi \left( x,t,\lambda \right) 
$ is eigenfunction corresponding to $\lambda $, and $L$ and $M$ are given in
the form 
\begin{eqnarray}
L &=&-i\lambda ({\bf S\cdot \sigma }),  \nonumber \\
M &=&\frac i2({\bf \varepsilon \cdot \sigma })+i2\lambda ^2({\bf S\cdot
\sigma })-\lambda ({\bf S\cdot \sigma })(\frac \partial {\partial x}{\bf %
S\cdot \sigma }).  \label{lax2}
\end{eqnarray}
Here ${\bf \sigma }$ is Pauli matrix. Thus Eq. (\ref{Eq2}) can be recovered
from the compatibility condition $\frac \partial {\partial t}L-\frac \partial
{\partial x}M+\left[ L,M\right] =0$. Based on the Lax equations (\ref{lax1}%
), we derive the exact $N$-soliton solution by employing the inverse
scattering transformation. We consider the following natural boundary
condition of initial time($t=0$), ${\bf S}\left( x\right) \equiv \left(
S^x,S^y,S^z\right) \rightarrow \left( \sin \theta ,0,\cos \theta \right) 
\text{ as}\left| x\right| \rightarrow \infty $, namely, the spin vector is
along the field direction. We then have the asymptotic form of Eq. (\ref
{lax1}) at $\left| x\right| \rightarrow \infty $, 
\begin{equation}
\partial _xE(x,\lambda )=L_0(\lambda )E(x,\lambda ),
\end{equation}
where 
\begin{equation}
E(x,\lambda )=Ue^{-i\lambda x\sigma _3},\ \ \ L_0(\lambda )=-i\lambda U_0,
\end{equation}
and 
\begin{equation}
U_0=\left( 
\begin{array}{cc}
\cos \theta & \sin \theta \\ 
\sin \theta & -\cos \theta
\end{array}
\right) ,U=\left( 
\begin{array}{cc}
1 & -\tan \frac \theta 2 \\ 
\tan \frac \theta 2 & 1
\end{array}
\right) .
\end{equation}
The Jost solutions $\Psi _{+}\left( x,\lambda \right) $ and $\Psi _{-}\left(
x,\lambda \right) $ of Eq. (\ref{lax1}) are defined as 
\begin{eqnarray}
\Psi _{+}(x,\lambda ) &\rightarrow &E\left( x,\lambda \right) \text{ \ \ \
as }x\rightarrow \infty ,  \nonumber \\
\Psi _{-}\left( x,\lambda \right) &\rightarrow &E\left( x,\lambda \right) 
\text{ \ \ \ as }x\rightarrow -\infty .  \nonumber
\end{eqnarray}
With standard procedures, one finds the following integral representations
of the Jost solutions in terms of the integration kernels $K$ and $N$ to be
determined, 
\begin{eqnarray}
&&\Psi _{+}(x,\lambda )=Ue^{-i\lambda x\sigma _3}+\lambda \int_x^\infty
dyK(x,y)Ue^{-i\lambda y\sigma _3},  \nonumber \\
&&K(x,\infty )=0,~K(x,y)=0~\text{as}~y<x.  \label{Int1}
\end{eqnarray}
and 
\begin{eqnarray}
&&\Psi _{-}\left( x,\lambda \right) =Ue^{-i\lambda x\sigma _3}+\lambda
\int_{-\infty }^xdyN\left( x,y\right) Ue^{-i\lambda y\sigma _3},  \nonumber
\\
&&N\left( x,-\infty \right) =0,N(x,y)=0~\text{as}~y<x.  \label{Int2}
\end{eqnarray}
where $K$ and $N$ are $2\times 2$ matrices. Substituting $\Psi _{+}\left(
x,\lambda \right) $ in Eq. (\ref{Int1}) into Eq.(\ref{lax1}) and noting $%
U\sigma _3U^{-1}=U_0$, we obtain 
\begin{equation}
{\bf S\cdot \sigma }=\left[ I-iK\left( x,x\right) U_0\right] U_0\left[
I-iK\left( x,x\right) U_0\right] ^{-1}  \label{Int3}
\end{equation}
where $I$ is unit matrix. It is obvious that Eq. (\ref{Int3}) gives rise to
a relation between kernel $K$ and spin vector ${\bf S}$ to be obtained.

The scattering data for the operator $L(x,\lambda )$ are the set $%
s=\{a(\lambda ),b(\lambda );\lambda _n,c_n,Im\lambda >0,n=1,\cdots ,N\}$,
where $|a(\lambda )|^2+|b(\lambda )|^2=1$, and the function $a(\lambda )$
can be analytically continued to the half-plane $Im\lambda >0$. The discrete
eigenvalues, $\lambda _n$, for the operator $L\left( x,\lambda \right) $ are
zeroes of $a(\lambda )$ such that $a(\lambda _n)=0$ (for the simplicity we
consider only simple zeroes). The functions $a(\lambda )$ and $b(\lambda )$
are seen to be transmission and reflection coefficients of the operator $L$
respectively. The parameter $c_n$ denotes the asymptotic characteristics of
the eigenfunctions.

The time-dependence of the scattering data $s\left( t\right) $ can be
obtained from the second Lax equation (\ref{lax1}), 
\begin{eqnarray}
a(\lambda ,t) &=&a(\lambda ,0),  \nonumber \\
b(\lambda ,t) &=&\exp \left( -4i\lambda ^2t-i\frac{g\mu _BB\sin \omega t}{%
J\omega }\right) b(\lambda ,0),  \nonumber \\
\lambda _n(t) &=&\lambda _n(0),  \nonumber \\
c_n(t) &=&\exp \left( -4i\lambda _n^2t-i\frac{g\mu _BB\sin \omega t}{J\omega 
}\right) c_n(0).
\end{eqnarray}
where $c_n(0)$, $b(\lambda ,0)$ and $a(\lambda ,0)$ are constants determined
by initial conditions. The Gelfand-Levitan-Marchenko equation establishes a
relation between the kernel $K\left( x,y,t\right) $ and the scattering data $%
s\left( t\right) $ and has the form 
\begin{equation}
K\left( x,y,t\right) U\left( 
\begin{array}{l}
1 \\ 
0
\end{array}
\right) +F_1+\frac 1{2\pi }\int_{-\infty }^\infty \lambda ^{-1}r\left(
\lambda \right) F_2d\lambda =0,  \label{ma1}
\end{equation}
as $y>x$, where $r\left( \lambda \right) =b\left( \lambda \right) /a\left(
\lambda \right) $ and 
\begin{eqnarray}
&&F_1=U\left( 
\begin{array}{l}
0 \\ 
1
\end{array}
\right) \sum_{n=1}^N\frac{c_n\left( t\right) }{\lambda _n}e^{i\lambda
_n\left( x+y\right) }  \nonumber \\
&&+\int_x^\infty K(x,z,t)U\left( 
\begin{array}{l}
0 \\ 
1
\end{array}
\right) \sum_{n=1}^Nc_n\left( t\right) e^{i\lambda _n\left( y+z\right) }dz, 
\nonumber \\
&&F_2=U\left( 
\begin{array}{l}
0 \\ 
1
\end{array}
\right) e^{i\lambda x}+\lambda \int_x^\infty K(x,z,t)U\left( 
\begin{array}{l}
0 \\ 
1
\end{array}
\right) e^{i\lambda z}dz.
\end{eqnarray}
For the reflectionless case, $r\left( \lambda \right) =0$, Eq. (\ref{ma1})
becomes a set of algebraic equations and after tedious calculation the
matrix elements of the kernel $K$ are obtained as 
\begin{eqnarray}
K_{11}(x,x,t) &=&\cos ^2\frac \theta 2[B_1+B_2\tan \frac \theta 2], 
\nonumber \\
K_{12}(x,x,t) &=&\cos ^2\frac \theta 2[B_1\tan \frac \theta 2-B_2].
\label{k1}
\end{eqnarray}
with 
\begin{eqnarray}
B_1 &=&\frac{\det [I+G^{\prime \prime }G^{\prime }+D^T(C\tan \frac \theta 2-%
\overline{C}G^{\prime })]}{\det (I+G^{\prime }G^{\prime \prime })}-1, 
\nonumber \\
B_2 &=&\frac{\det [I+G^{\prime }G^{\prime \prime }-\overline{D}^T(\overline{C%
}+CG^{\prime \prime }\tan \frac \theta 2)]}{\det (I+G^{\prime }G^{\prime
\prime })}-1.
\end{eqnarray}
where $C\left( x,t\right) $, $C^{\prime }\left( x,t\right) $, $D\left(
x\right) $ are $1\times N$ matrices, $G^{\prime }\left( x,t\right) $, $%
G^{\prime \prime }\left( x,t\right) $ $N\times N$ matrices, respectively.
The superscript $T$ means the transposed matrix and the overbar denotes
complex conjugate, 
\begin{eqnarray}
C(x,t)_n &=&c_n\left( t\right) \lambda _n^{-1}D\left( x\right) _n,  \nonumber
\\
C^{\prime }\left( x,t\right) _n &=&c_n\left( t\right) D\left( x\right) _n, 
\nonumber \\
D(x)_n &=&\exp (i\lambda _nx),  \nonumber \\
G^{\prime }\left( x,t\right) _{nm} &=&\frac 1{i\left( \overline{\lambda }%
_n-\lambda _m\right) }\overline{D\left( x\right) }_nC^{\prime }\left(
x,t\right) _m,  \nonumber \\
G^{\prime \prime }\left( x,t\right) _{nm} &=&\frac 1{-i\left( \lambda _n-%
\overline{\lambda }_m\right) }D\left( x\right) _n\overline{C^{\prime }\left(
x,t\right) }_m.
\end{eqnarray}
Substituting Eq. (\ref{k1}) into Eq. (\ref{Int3}) , we obtain the general
form of N-soliton trains, 
\begin{eqnarray}
S^x &=&\frac 1\Delta 
\mathop{\rm Re}%
\left\{ {}\right. -i2K_{12}\left[ 1-iK_{11}\cos \theta \right]  \nonumber \\
&&+[1+K_{11}^2-K_{12}^2]\sin \theta \left. {}\right\} ,  \nonumber \\
S^y &=&\frac{-1}\Delta 
\mathop{\rm Im}%
\left\{ {}\right. -i2K_{12}\left[ 1-iK_{11}\cos \theta \right]  \nonumber \\
&&+[1+K_{11}^2-K_{12}^2]\sin \theta \left. {}\right\} ,  \nonumber \\
S^z &=&\frac 1\Delta \left\{ {}\right. [1+\left| K_{11}\right| ^2-\left|
K_{12}\right| ^2]\cos \theta  \nonumber \\
&&+2%
\mathop{\rm Im}%
\left[ K_{11}\left( 1+i\overline{K_{12}}\sin \theta \right) \right] \left.
{}\right\} .  \label{nsoliton}
\end{eqnarray}
where $\overline{K_{12}}$ is the complex conjugate of $K_{12}$. 
\begin{eqnarray}
\Delta &=&\left| 1-i\left[ K_{11}\cos \theta +K_{12}\sin \theta \right]
\right| ^2  \nonumber \\
&&+\left| K_{11}\sin \theta -K_{12}\cos \theta \right| ^2.
\end{eqnarray}
According to exact N-soliton solutions in Eq. (\ref{nsoliton}), we,
generally speaking, can investigate the dynamics of soliton trains and
soliton interaction. The neighboring solitons may repulse or attract each
other with a force depending on their phase difference. Particularly we in
the following shall concentrate on the analyses of one-soliton dynamics and
two-soliton collisions which may be of more interest.

\section{One-soliton dynamics and Spin precession with time varying amplitude
}

When $N=1$, from Eq. (\ref{k1}) and Eq. (\ref{nsoliton}) we obtain the
general form of the exact one-soliton solution as follows 
\begin{eqnarray}
S^x &=&\frac{R_1}{|\lambda _1|^4\cosh ^2\Theta _1},  \nonumber \\
S^y &=&\frac{R_2}{|\lambda _1|^4\cosh ^2\Theta _1},  \nonumber \\
S^z &=&R_3\cos \theta +R_4\sin \theta ,  \label{onesoliton1}
\end{eqnarray}
where 
\begin{eqnarray}
R_1 &=&[|\lambda _1|^4\cosh ^2\Theta _1+\beta _1^2(\alpha _1^2-\beta
_1^2\cos 2\theta )e^{-2\Theta _1}]\sin \theta  \nonumber \\
&&+\beta _1^2|\lambda _1|^2(2\cos ^2\theta \sin ^2\Phi _1-1)\sin \theta 
\nonumber \\
&&+2\beta _1^2|\lambda _1|(2\beta _1\sin \Phi _1\sin ^2\theta +\alpha _1\cos
\Phi _1)e^{-\Theta _1}\cos \theta  \nonumber \\
&&-2\beta _1(\beta _1e^{-\Theta _1}\sin \theta +|\lambda _1|\sin \Phi _1\cos
\theta )[|\lambda _1|^2\cosh \Theta _1  \nonumber \\
&&+\beta _1(|\lambda _1|\sin \Phi _1\sin \theta -\beta _1e^{-\Theta _1}\cos
\theta )\cos \theta ],  \nonumber \\
R_2 &=&2\alpha _1\beta _1^2|\lambda _1|e^{-\Theta _1}\sin \Phi _1+2\beta
_1|\lambda _1|^3\cos \Phi _1\cosh \Theta _1  \nonumber \\
&&-2\beta _1^3|\lambda _1|\left( \sin \theta +\cos ^2\theta \right)
e^{-\Theta _1}\cos \Phi _1,  \nonumber \\
R_3 &=&1-\frac{2\beta _1^2}{|\lambda _1|^2\cosh ^2\Theta _1},  \nonumber \\
R_4 &=&\frac 1{|\lambda _1|^2\cosh ^2\Theta _1}\left[ {}\right. 2\beta
_1^2\cos (\Phi _1-\phi _1)\sinh \Theta _1  \nonumber \\
&&+2\alpha _1\beta _1\sin (\Phi _1-\phi _1)\cosh \Theta _1\left. {}\right]
\label{onesoliton2}
\end{eqnarray}
with 
\begin{eqnarray}
\Theta _1 &=&2\beta _1(x-V_1t)-x_1,  \nonumber \\
V_1 &=&4\alpha _1,x_1=\ln [(2\beta _1)^{-1}c_1\left( 0\right) ],  \nonumber
\\
\Phi _1 &=&2\alpha _1x-4(\alpha _1^2-\beta _1^2)t-\left( \omega J\right)
^{-1}g\mu _BB\sin (\omega t)-\phi _1,  \nonumber \\
\Omega _1 &=&2\alpha _1^{-1}\left( \alpha _1^2-\beta _1^2\right) +\Omega _B,
\nonumber \\
\Omega _B &=&\left( 2\alpha _1J\right) ^{-1}g\mu _BB\cos \omega t,
\label{onesoliton3}
\end{eqnarray}
$\phi _1=\arg \lambda _1$, $\lambda _1=\alpha _1+i\beta _1$ is eigenvalue
parameter. The solution (\ref{onesoliton1}) describes a spin precession
around magnetic field direction characterized by four real parameters:
velocity $V_1$, frequency $\Omega _1$, coordinate of the center of the
solitary wave $x_1$ and initial phase $\phi _1$. The center of solitary wave
moves with a velocity $V_1$, while the wave depth and width vary
periodically with time. The wave shape is modulated periodically by
frequency $\Omega _1$ depending on magnetic field. Therefore, the solution (%
\ref{onesoliton1}) can not be written as the form of separating variables.
Amplitude $A$ and phase $\Phi _1$ are complicated functions of $J$, $B$, $%
\omega $ and $\lambda _1$. When $\alpha _1=\beta _1$, the frequency $\Omega
_1$ depends on magnetic field only, and we have $\Omega _1=\Omega _B$. If $%
\alpha _1=\beta _1$ and $B=0$, the solution (\ref{onesoliton1}) reduces to
the usual soliton without shape changing. Therefore, we can use magnetic
field to adjust spin precession and the wave shape as well.

For a special case, $\theta =0$, namely the magnetic field is along the
z-axis, $S^z$ is independent of magnetic field, $S^z=R_3$, while $S^x$ and $%
S^y$ precess around magnetic field (z-axis). The precession frequency $%
\Omega _1$ is determined by magnetic field. As magnetic field rotates from $%
\theta =0$ (z-axis) to $\theta =\pi /2$ (x-axis), we can find the
correspondence such that $S^x\rightarrow -S^z$, $S^y\rightarrow S^y$, $%
S^z\rightarrow S^x$. The three components of spin vector satisfy ``left-hand
rule''. When $\theta =\pi /2$, $S^x$ is independent of magnetic field, while 
$S^y$ and $S^z$ precess around magnetic field (x-axis). These results show
that the magnetic field results in the motion of the center of solitary
waves along the field direction and the spin vector rotates around the field
in any case.

\section{Two-soliton collision}

When $N=2$, from Eq. (\ref{k1}) and Eq. (\ref{nsoliton}) the general form of
the exact two-soliton solution is seen to be 
\begin{eqnarray}
S^x &=&%
\mathop{\rm Re}%
[-i2Q_2(1-iQ_1\cos \theta )+(1+Q_1^2-Q_2^2)\sin \theta ],  \nonumber \\
S^y &=&%
\mathop{\rm Im}%
[i2Q_2(1-iQ_1\cos \theta )-(1+Q_1^2-Q_2^2)\sin \theta ],  \nonumber \\
S^z &=&(1+\left| Q_1\right| ^2-\left| Q_2\right| ^2)\cos \theta  \nonumber \\
&&+2%
\mathop{\rm Im}%
[Q_1(1+i\overline{Q_2}\sin \theta )].  \label{twosoliton1}
\end{eqnarray}
where 
\begin{eqnarray}
Q_1 &=&\frac{\cos ^2\frac \theta 2}W\{\left( f_1-f_3\right) f_6+\left(
f_2-f_4\right) f_5  \nonumber \\
&&+\tan \frac \theta 2[(\overline{f_1}-\overline{f_3})f_8+(\overline{f_2}-%
\overline{f_4})f_7]\},  \nonumber \\
Q_2 &=&\frac{\cos ^2\frac \theta 2}W\{[\left( f_1-f_3\right) f_6+\left(
f_2-f_4\right) f_5]\tan \frac \theta 2  \nonumber \\
&&-(\overline{f_1}-\overline{f_3})f_8-(\overline{f_2}-\overline{f_4})f_7\},
\end{eqnarray}
with 
\begin{eqnarray}
f_1 &=&1+\left| q_1\right| ^2+\chi _1\overline{\chi }_2q_1\overline{q}_2,%
\text{ }f_2=1+\left| q_2\right| ^2+\overline{\chi }_1\chi _2\overline{q}%
_1q_2,  \nonumber \\
f_3 &=&\overline{\chi }_1\left| q_1\right| ^2+\chi _1q_1\overline{q}_2,\text{
}f_4=\overline{\chi }_2\left| q_2\right| ^2+\chi _2\overline{q}_1q_2, 
\nonumber \\
f_5 &=&\xi _1\left( q_1\tan \frac \theta 2-\left| q_1\right| ^2\right) -\chi
_1\xi _2q_1\overline{q}_2,  \nonumber \\
f_6 &=&\xi _2\left( q_2\tan \frac \theta 2-\left| q_2\right| ^2\right) -\chi
_2\xi _1\overline{q}_1q_2,  \nonumber \\
f_7 &=&-\xi _1\left( \overline{q}_1+\left| q_1\right| ^2\tan \frac \theta 2%
\right) -\xi _2\overline{\chi }_1\overline{q}_1q_2\tan \frac \theta 2, 
\nonumber \\
f_8 &=&-\xi _2\left( \overline{q}_2+\left| q_2\right| ^2\tan \frac \theta 2%
\right) -\xi _1\overline{\chi }_2q_1\overline{q}_2\tan \frac \theta 2, 
\nonumber \\
\chi _1 &=&\frac{2\beta _1\lambda _1}{-i\left( \lambda _1-\overline{\lambda }%
_2\right) \left| \lambda _1\right| },\chi _2=\frac{2\beta _2\lambda _2}{%
-i\left( \lambda _2-\overline{\lambda }_1\right) \left| \lambda _2\right| },
\nonumber \\
W &=&f_1f_2-f_3f_4,\text{ }q_j=e^{-\Theta _j+i\Phi _j},\text{ }\xi _j=2\beta
_j\left| \lambda _j\right| ^{-1},
\end{eqnarray}
and 
\begin{eqnarray}
\Theta _j &=&2\beta _j(x-V_jt)-x_j,  \nonumber \\
V_j &=&4\alpha _j,x_j=\ln [(2\beta _j)^{-1}c_j\left( 0\right) ],  \nonumber
\\
\Phi _j &=&2\alpha _jx-4(\alpha _j^2-\beta _j^2)t-\left( \omega J\right)
^{-1}g\mu _BB\sin (\omega t)-\phi _j,  \nonumber \\
\Omega _j &=&2\alpha _j^{-1}\left( \alpha _j^2-\beta _j^2\right) +\Omega _B,
\nonumber \\
\Omega _B &=&\left( 2\alpha _jJ\right) ^{-1}g\mu _BB\cos \omega t,
\end{eqnarray}
here $\phi _j=\arg \lambda _j$ and $\lambda _j=\alpha _j+i\beta _j$ is
eigenvalue parameter, $j=1,2$. The solutions (\ref{twosoliton1}) describe a
general inelastic scattering process of two solitary waves with different
center velocities $V_1$ and $V_2$, different shape variation frequencies $%
\Omega _1$ and $\Omega _2$. Before collision, they move towards each other,
one with velocity $V_1$ and shape variation frequency $\Omega _1$, the other
with $V_2$ and $\Omega _2$. The interaction potential between two solitons
is a complicated function of parameters $J,B,\omega $ and $\lambda _j$. When 
$\alpha _j=\beta _j$, two-soliton shape-variation frequencies $\Omega
_j(j=1,2)$ are determined by magnetic field. In the case of $B=0$, the
solutions (\ref{twosoliton1}) reduce to that of the usual two-soliton with
two center velocities while without shape change where a interesting process
in the absence of magnetic field is that the collision can result in the
interchange of amplitude $A_j$ and phase $\Phi _j(j=1,2)$ like exactly in
the case of elastic collision of two particles.

In order to understand the nature of two-soliton interaction, we analyze
asymptotic behavior of two-soliton solutions (\ref{twosoliton1}).
Asymptotically, the two-soliton waves (\ref{twosoliton1}) can be written as
a combination of two one-soliton waves (\ref{onesoliton1}) with different
amplitude and phase. The formation of two-soliton waves in the corresponding
limits $x\rightarrow -\infty $ and $x\rightarrow \infty $ is similar to that
of one-soliton waves (\ref{onesoliton1}). Analysis reveals that there is an
amplitude exchange among three components $S^x$, $S^y$ and $S^z$ of each
soliton during collision, which can be described by a transition matrix $%
T_l^k$ such that $A_l^{k+}=A_l^{k-}T_l^k$, where the subscript $l=1,2$
respectively represents the first and the second soliton, $k=x,y,z$ denote
three components of each soliton, the sign $\pm $ denotes the asymptotic
limits of the corresponding amplitude, $A_l^{k\pm }$, at $x\rightarrow \pm
\infty $. As a consequence, amplitude change of the three components $S_1^k$
of the first soliton from $A_1^{k-}$ to $A_1^{k+}$ is given by square of
transition matrices $|T_1^k|^2$ along with phase shift $\delta \Phi _1^k$
during collision. In a similar fashion, the three components $S_2^k$ of the
second soliton also change amplitudes from $A_2^{k-}$ to $A_2^{k+}$ with a
quantity $|T_2^k|^2$. The associate phase shift for the second soliton is $%
\delta \Phi _2^k$. We also note a net change of separation distance between
two solitons by $\delta X_{12}$.

For the special case $|T_l^k|=1$, which is possible only when $\lambda _2=-%
\overline{\lambda }_1$, we have the standard elastic collision. For all
other cases, we have the quantity $|T_l^k|\not{=}1$, which corresponds to
relative change among three components of the spin vector leading to the
deformation of soliton shape. However, the total amplitude of individual
solitons $S_1$ and $S_2$ is conserved quantity i.e., $\sum_l|A_l^{k\pm }|^2$
is constant for $l=1,2$.

It is interesting to show the inelastic collision graphically. The general
inelastic head on collision is explained in Fig. 1 from which it is seen
that the amplitudes of $S_1$, $S_2$ are respectively suppressed and enhanced
after collision. Fig. 2 are devoted to the complete inelastic head on
collisions. The amplitudes of $S_1$and $S_2$ are respectively suppressed
after collision shown in Fig. 2a and 2b. The complete inelastic
overtake-collision is shown in Fig. 3 with the amplitudes of $S_1$and $S_2$
suppressed, respectively.

\section{Conclusion}

In terms of an inverse scattering transformation the exact solution of
N-soliton trains in a spin chain driven by a time oscillating magnetic field
is obtained. From the general solution the dynamics and soliton interactions
are analyzed. The one-soliton solution gives rise explicitly to the spin
precession along with the soliton shape variation induced by the time
varying field. It is also shown that the time varying field leads generally
to the inelastic and particularly the complete inelastic two-soliton
collisions which may be useful in developing a soliton-shape control
technique.

\section{Acknowledgment}

This work was supported by the NSF of China under Grant Nos. 10194095,
90103024 and 10075032.

Figure caption

Fig. 1

Inelastic head on collision between two solitons -- profiles of z-component $%
S^z(x;t)$ of spin vector in Eq. (22) in spin chain under a time-dependent
magnetic field showing two different dramatic scenarios of the shape
changing collision, where $\theta =\frac \pi {36}$, $\lambda _1=-0.2+i0.45$, 
$\lambda _2=0.3+i0.65$, $c_1\left( 0\right) =-0.2$, $c_2\left( 0\right) =3.5$%
, $g\mu _BB/J=0.01$, $\omega =10$, $V_1=-0.8$, $V_2=1.2$. All quantities
plotted are dimensionless. The same is in Fig. 2 and 3.

Fig. 2

(2a) Complete inelastic head on collision expressed by Eq. (22) when $S_1$
suppressed, where $\theta =0$, $\lambda _1=-0.35+i0.4$, $\lambda _2=0.2-i0.6$%
, $c_1\left( 0\right) =0.2$, $c_2\left( 0\right) =-2.5$, $g\mu _BB/J=0.01$, $%
\omega =10$, $V_1=-1.4$, $V_2=0.8$.

(2b) Complete inelastic head on collision expressed by Eq. (22) when $S_2$
suppressed, where $\theta =0$, $\lambda _1=-0.35-i0.4$, $\lambda _2=0.2+i0.6$%
, $c_1\left( 0\right) =-0.2$, $c_2\left( 0\right) =2.5$, $g\mu _BB/J=0.01$, $%
\omega =10$, $V_1=-1.4$, $V_2=0.8$.

Fig. 3

(3a) Complete inelastic overtake-collision expressed by Eq. (22) when $S_1$
suppressed, where $\theta =0$, $\lambda _1=-0.55+i0.4$, $\lambda
_2=-0.1-i0.45$, $c_1\left( 0\right) =0.2$, $c_2\left( 0\right) =-2.5$, $g\mu
_BB/J=0.01$, $\omega =10$, $V_1=-2.2$, $V_2=-0.4$.

(3b) Complete inelastic overtake-collision expressed by Eq. (22) when $S_1$
suppressed, where $\theta =0$, $\lambda _1=-0.55-i0.4$, $\lambda
_2=-0.1+i0.45$, $c_1\left( 0\right) =-0.2$, $c_2\left( 0\right) =2.5$, $g\mu
_BB/J=0.01$, $\omega =10$,$V_1=-2.2$, $V_2=-0.4$.

\end{document}